\documentclass[mathleft]{an}
\usepackage{graphicx}
\usepackage{times}
\def\gtsima{$\; \buildrel > \over \sim \;$}
\def\ltsima{$\; \buildrel < \over \sim \;$}
\def\gsim{\lower.5ex\hbox{\gtsima}}
\def\lsim{\lower.5ex\hbox{\ltsima}}

\def\Lya{Ly$\alpha$~}

\newcommand{\CIV}{\mbox{C\,{\sc iv}}}

\newcommand{\CII}{\mbox{C\,{\sc ii}}}

\newcommand{\OI}{\mbox{O\,{\sc i}}}
\newcommand{\MgI}{\mbox{Mg\,{\sc i}}}
\newcommand{\MgII}{\mbox{Mg\,{\sc ii}}}
\newcommand{\AlII}{\mbox{Al\,{\sc ii}}}

\newcommand{\SiII}{\mbox{Si\,{\sc ii}}}

\newcommand{\FeII}{\mbox{Fe\,{\sc ii}}}
\newcommand{\ZnII}{\mbox{Zn\,{\sc ii}}}
\newcommand{\HI}{\mbox{H\,{\sc i}}}
\newcommand{\HII}{\mbox{H\,{\sc ii}}}

\begin{document}

\Pagespan{789}{}
\Yearpublication{}%
\Yearsubmission{}%
\Month{}%
\Volume{}%
\Issue{}%

\title{Optical-NIR spectra of quasars close to reionization ($z\sim
  6$)\thanks{Based on observations collected at the European Southern
    Observatory Very Large Telescope, Cerro Paranal, Chile -- Programs
    084.A-0550(A) and   085.A-0299(A)
}}

\author{V. D'Odorico\inst{1}\fnmsep\thanks{Corresponding author:
  \email{dodorico@oats.inaf.it}\newline}
\and G. Cupani\inst{1}
\and S. Cristiani\inst{1},\inst{2}
\and R. Maiolino\inst{3}
\and P. Molaro\inst{1}
\and M. Nonino\inst{1}
\and A. Cimatti\inst{4}
\and S. di Serego Alighieri\inst{5}
\and F. Fiore\inst{3}
\and A. Fontana\inst{3}
\and S. Gallerani\inst{3}
\and E. Giallongo\inst{3}
\and F. Mannucci\inst{5}
\and A. Marconi\inst{6}
\and L. Pentericci\inst{3}
\and M. Viel\inst{1},\inst{2}
\and G. Vladilo\inst{1}
}
\titlerunning{Optical-NIR spectra of quasars at $z\sim6$}
\authorrunning{V. D'Odorico et al.}
\institute{INAF-OATS, Via Tiepolo 11, 34143 Trieste, Italy 
\and 
INFN/National Institute of Nuclear Physics, via Valerio 2, 34127
Trieste, Italy
\and
INAF-OAR, via di Frascati 33, 00040 Monte Porzio Catone, Italy
\and
Dipartimento di Astronomia, Universit\`a di Bologna Via Ranzani 1, I-40127 Bologna, Italy
\and
INAF-OAArcetri, Largo Enrico Fermi 5, 50125 Firenze. Italy
\and
Dipartimento di Fisica e Astronomia, Universit\`a di Firenze, Largo E. Fermi 2, Firenze, Italy
}

\received{}
\accepted{}
\publonline{later}


\abstract{X-shooter, with its characteristics of resolution, spectral coverage
and efficiency, provides a unique opportunity to obtain spectra of the
highest-redshift quasars ($z\sim6$) that will allow us to carry out
successful investigations on key cosmological issues, from the details
of the re-ionization process, to the evolution of the first galaxies
and AGNs. In this paper, we present the spectra of three $z\sim6$ quasars:
one obtained during the commissioning of X-shooter and two in the
context of our ongoing GTO programme. Combining this sample with data in the literature, we update the value of the \CIV\  cosmic mass density in the range $4.5 \le z \le 5$, confirming the constant trend with redshift between 2.5 and 5.    }

\maketitle

\section{Introduction}
The properties of stars, galaxies and quasars in the local and early
Universe can be investigated through their impact on the intergalactic
medium (IGM). In particular, the radiation emitted and the metals
ejected from these objects re-ionized and polluted the IGM. 
As a consequence, the detailed understanding of these mechanisms has
the potential to significantly constrain models for the formation and
evolution of galaxies and quasars, and the re-ionization history of the
Universe.
The IGM is mainly studied through the absorption signature it leaves
in bright high-redshift sources, quasars and Gamma-Ray Bursts
(GRBs). The highest redshift quasars have been detected mainly by the
Sloan Survey (SDSS) at $z\sim6$ (e.g. Fan et al. 2001), corresponding
to $\sim1$ billion years after the Big Bang. 
This sample of $\sim 20$ objects has been used to investigate several
topics, in particular the ionization and chemical status of the IGM at
these high redshifts.
 
The fact that the spectra of the high-$z$ quasars show some extended
regions of zero transmission in the \Lya\ forest (e.g. Becker et
al. 2001) was interpreted as the signature of relatively large
\HI\ densities at those redshifts (the so called ``Gunn-Peterson
effect'', Gunn \& Peterson 1965).  
However, due to the high sensitivity of the Gunn-Peterson
(GP) optical depth, $\tau_{\rm GP}$, to tiny neutral hydrogen fractions
($x_{\rm HI}>10^{-3} - 10^{-4}$), the detection of a GP trough only
translates into a lower limit for the volume averaged neutral hydrogen
fraction. Nevertheless, the sudden rise of the \Lya\ opacity 
approaching $z\sim6$, was considered as an evidence for the
reionization of the Universe to be completed by this epoch (e.g., Fan
et al. 2006, but see also Becker et al. 2007).
More sophisticated approaches to the investigation of the reionization
epoch with the \Lya\ forest are represented by the statistics of the
dark gaps (e.g., Fan et al. 2006; Gallerani et al. 2006, 2008) and the
application of the {\sl apparent shrinking criterion} (Maselli et
al. 2007, 2009) which compares the observed size of the flux transmission window around the quasar with the size of the quasar \HII\ region  predicted by radiative transfer simulations. 
Both methods applied to the sample of spectra available from the
literature  suggest that the IGM at $z\sim6$ is already highly ionized. However,
systematic errors (due e.g. to the uncertainty in the systemic
redshift of the quasar or in the position of the level of the
continuum) still play an important role and undermine the reliability of the obtained results.

The same high-$z$ quasars can be used to put an indirect constraint on
the epoch of re-ionization by investigating the redshift evolution of
metal abundances traced by ionic 
absorption lines. Indeed, the main sources of ionizing photons are now
thought to be massive stars, which are also the bulk producers of
heavy elements. 
The investigation of the regime beyond $z\sim5$ is essential since in this
redshift range the comoving star formation rate density appears to
decline (e.g. Mannucci et al. 2007; Gonzalez et al 2010). If a
similar behaviour is observed for the mass density of metals in the
IGM, then a scenario where winds from massive star-forming galaxies
pollute the IGM with metals would be favoured.  On the other hand, if
the mass density of metals is observed to remain constant, this would 
point to an epoch of very early enrichment of the IGM, presumably by
smaller mass objects (e.g. PopIII stars), when shallow potential wells
allow winds to distribute metals over large comoving volumes, thus
producing a quite uniform metallicity distribution (e.g. Madau et
al. 2001). 
The most recent compilation of the evolution with redshift of the
\CIV\ cosmic mass density (D'Odorico et al. 2010) shows a flat
behaviour in the range $z\simeq 3-5$ and a possible downturn at
$z>5$. The latter result needs confirmation since the point at $z\simeq5.5$
is based on just four \CIV\ absorptions (Ryan-Weber et al. 2009). The
more conservative result by Becker et al. (2009) is consistent with an
invariant value up to $z\sim6$. 
Intergalactic absorption-line detections at these redshifts are rare
because the \CIV\ doublet moves into the near-infrared spectral
region.  Absorption line spectroscopy is much more challenging in this
regime because of increased detector noise, OH emission from the sky
and more severe telluric absorption. 

As soon as metals start to be produced and ejected in the diffuse
medium, part of them condenses into microscopic solid particles,
giving birth to the interstellar dust of the first
galaxies. Characterizing the properties of this dust at high-$z$ is 
important to improve our understanding of the early cosmic epochs. 
High redshift quasars contain large dust masses, which have been
revealed by mm and sub-mm observations
Yet, the origin of dust at such early epochs is still
unclear. Among the stellar sources, the cool and dense atmosphere 
of Asymptotic Giant Branch (AGB) stars and the expanding ejecta of
core collapse supernovae (SN) offer the most viable sites of dust
grains condensation. However, AGB stars require about 1 Gyr to evolve
in large number, and therefore to effectively enrich the interstellar
medium with dust (but see Valiante et al. 2009). This timescale is comparable to the age of the
Universe at redshift  $z\sim6$. 
On the other hand, since dust pollution by SNe can occur on shorter timescales ($\sim 10^6$ yr), a SN origin has often been advocated as the only possible explanation for the large amount of dust observed in high redshift quasars.
This scenario has been tested
through observations of the reddened quasar SDSS J1048+4637 at $z_{\rm em}=6.2$
(Maiolino et al. 2004), the $z_{\rm em}=6.29$ GRB 050904 (Stratta et al. 2007) 
and the $z_{\rm em}\simeq5$ GRB 071025 (Perley et al. 2010). In these sources,
the inferred dust extinction curve is different with respect to any of
the extinction curves observed at low $z$, and it shows a very good
agreement with the extinction curve predicted for dust formed in SN
ejecta. This is an indication that the properties of dust may evolve
beyond $z>5-6$. Recently, Gallerani et al. (2010) have analyzed the
dust properties of a sample of 33 quasars with $4<z_{\rm em}<6.4$. The 7 
quasars in the sample showing significant reddening are well described
by SN-type extinction curves. Marginal evidence of reddening is
detected in almost all the quasars of the sample, however a conclusive
result cannot be reached due to the limited resolution ($R < 800$) and 
wavelength coverage of the available spectra.

The X-shooter spectrograph, with its high sensitivity, extended
spectral coverage and intermediate resolution (see contributions by
S. D'Odorico, J. Vernet and F. Zerbi), appears to be the ideal
instrument to obtain data of the needed quality to allow significant
steps forward in the above fields of research.  
For this reason, we have proposed a programme (P.I. V. D'Odorico) to
observe the brightest (J$_{\rm Vega}\lsim 19$) quasars known with $z_{\rm em}\gsim5.7$ 
observable from Paranal, in the Italian guaranteed time of observation
(GTO).
7 objects satisfy our selection criteria. In the meantime, two of them
(SDSS J1306+0356 and ULAS J1319+0950) have been extensively observed by a
competing programme carried out in open time. 

In the following, we describe the present status of our programme.   

\section{Data observation, reduction and analysis}

The total time assigned to our GTO programme was of $\sim6$ nights (52
hours) and observations started in January 2010. GTO is carried out in
visitor mode. Due to bad weather and technical problems, we have lost more
than half of the assigned nighttime, up to now. 
At present, we have obtained the spectra of two quasars: J0818+1722 and
J1509-1749 which are described below, together with the spectrum of
J1306+0356 observed during the third commissioning run. The details of
the observations are reported in Table~\ref{tab_obs}. 

In the 2.5 nights still available to complete the programme, we plan
to observe three more objects.   

\begin{table}
\begin{center}
\caption{Journal of observations. All spectra were obtained using the
  combination of slits 1.0/0.9/0.9 arcsec in the UVB/VIS/NIR arms,
  respectively. }
\begin{minipage}{80mm}
\label{tab_obs}
\begin{tabular}{@{} l l l c r }
\hline  
QSO & $z_{\rm em}$ & Jmag & Date (UT) & Exp. (s)  \\
\hline
 J1306+0356$^1$ & 6.016 & 18.77 & 2009 Mar 19 & $5400$ \\
 J0818+1722$^2$ & 6.00 & 18.54 & 2010 Jan 21 & $19600$   \\
 &  & & 2010 Jan 22 & $3310$   \\
 J1509-1749$^3$ & 6.12 & 18.78 & 2010 May 18 & $18000$  \\
 &  &  & 2010 May 19 & $6000$  \\
\hline
\end{tabular}

(1) Fan et al. 2001; (2) Fan et al. 2006; (3) Willott et al. 2007
\end{minipage}
\end{center}
\end{table}

All the raw frames were reduced with the public release of the
X-shooter pipeline (see contributions by P. Goldoni and
by A. Modigliani). 
We followed the standard steps of the reduction, except for the
extraction of the 1D spectra from the 2D merged spectra for which we
used the command {\it extract/long} in MIDAS (using a predefined aperture)
which gives better results than the  pipeline, at least for such faint
objects. 
The instrument response curve was obtained reducing with the specific
pipeline recipe the standard flux stars observed the same night of the
scientific observations. 
Each extracted frame was then flux calibrated by dividing for the
response curve.  
Finally, the set of 1D spectra obtained for each object was added with a
weighted sum to obtain the final spectrum. 

The continuum level in the region redward of the \Lya\ emission was determined interpolating with a spline polynomial of 3rd degree the portions of the spectrum free from absorption lines. The same approach cannot be applied to the heavily absorbed \Lya\ forest. The continuum in this spectral range was obtained by extrapolating the power law which fits the red region cleaned from the intrinsic emission lines.  

Finally, the VIS and NIR spectra were corrected for telluric absorption dividing by the normalized spectrum of  standard spectroscopic stars
observed with the same instrumental set-up as the QSOs in our sample, using the command {\it telluric} in IRAF.  

In the following, all the reported signal to noise ratios (SNR) are per wavelength bin:
0.4 \AA\ in the VIS arm and 0.6 \AA\ in the NIR arm. 

\subsection{SDSS J1306+0356}

This object was studied in the past at low/intermediate resolution in
particular with GNIRS at Gemini South Observatory and with ISAAC at
the VLT. 
%
The spectrum analysed here was obtained during the 3rd commissioning of
X-shooter and has a relatively low SNR with respect 
to the other two spectra: $\simeq 30$ on the \Lya\ emission,
decreasing to 8-10 down to 1 $\mu$m.  
 
\begin{figure}
\includegraphics[width=8cm,height=8cm]{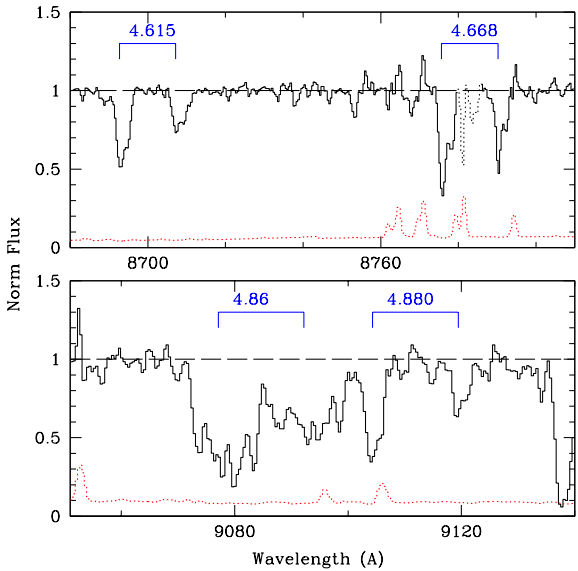}
\caption{\CIV\ absorptions detected in the spectrum of J1306+0356. The
doublets are marked together with their absorption redshift.}
\label{CIV_J1306}
\end{figure}

\begin{figure}
\includegraphics[width=8cm,height=8cm]{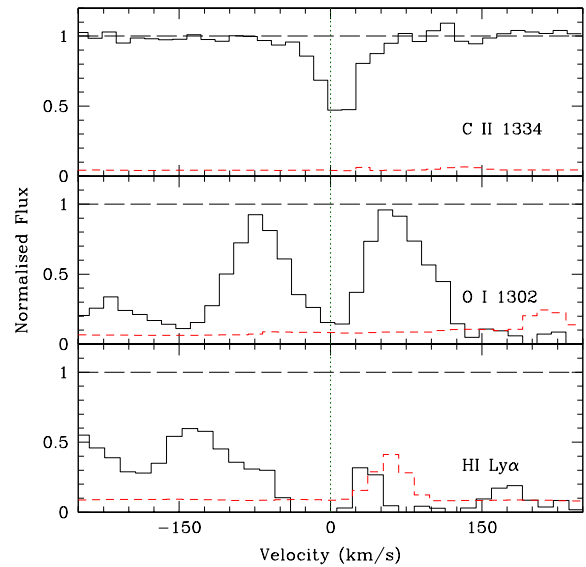}
\caption{Possible absorption system at $z_{\rm abs}=5.4347$ in the
  spectrum at J1306+0356. The \OI\ line falls in the \Lya\ forest.}
\label{z5p43_J1306}
\end{figure}

Jiang et al. (2007) detected 4 \MgII\ systems at $z_{\rm abs} = 2.20$, 2.53,
4.86 and 4.88 (see also Kurk et al. 2007).  
We confirm the presence of the three higher-redshift \MgII\ systems. On the other hand,  we identify the two lines previously interpreted as the
\MgII\ doublet at $z_{\rm abs}=2.20$ as the \SiII\ 1526
\AA\ absorptions at $z_{\rm abs}=4.86$ and 4.88, 
respectively. For these two systems we detect also the \CIV\ transitions
(see Fig.~\ref{CIV_J1306}). 
The line identified as \MgI\  2852 for the system at $z_{\rm abs}=2.20$
is instead \FeII\   2586 at $z_{\rm abs}=2.53$. 

We detect two more \CIV\ systems at $z_{\rm abs}=4.615$ (with associated
\SiII\ 1526 and \AlII\ 1670) and 4.668, a
possible weak \MgII\ doublet at $z_{\rm abs}=2.309$ and a possible system at
$z_{\rm abs}=5.435$ with \OI\ 1302, and
\CII\ 1334 \AA\ (see Fig.~\ref{z5p43_J1306}). 

Previous observations did not identify reliable \CIV\ absorptions in
the Z, J bands (corresponding to $z_{\rm abs} \simeq 5-6$, see Ryan-Weber
et al. 2006; Simcoe 2006), and 
unfortunately the NIR portion of the X-shooter spectrum of J1306+0356 has a
SNR too low to be used. 

\subsection{SDSS J0818+1722}

The spectrum of J0818+1722 is the best of our present sample, with a
SNR varying between $\sim 80$ and 20 in the VIS region (redward of the
\Lya\ emission) and 17 and 45 in the \CIV\ forest extending into the
NIR region.  
Thanks to the high SNR, we could subtract quite well the telluric
features, in particular in the VIS region, revealing the presence of
many  absorption systems.
 In particular, we detected \CIV\ doublets at
$z_{\rm abs}=4.463$, 4.498, 4.508, 4.523, 4.552, 4.577, 4.620,
4.727(complex), 4.732, 4.877, 4.942, 5.076, 5.308, 5.322,
5.324; two possible, very weak \MgII\ doublets at $z_{\rm abs}=2.0906$ and 2.129 and one very well defined with associated \FeII\ lines at $z_{\rm abs}=3.5628$ (in the NIR portion of the spectrum) . 

Previous spectroscopic studies of this SDSS QSO have been carried out
with NIRSPEC at the Keck telescope (Becker et al. 2009) and with ISAAC at
the VLT (Ryan-Weber et al. 2009). The latter spectrum shows a tentative
\CIV\ doublet at $z_{\rm abs}=5.7899$ and an \MgII\ doublet at $z_{\rm abs}=2.8341$. 
The presence of the \MgII\ system is not confirmed by our spectrum,
 we could not find any line at this redshift, neither the
\MgII\ doublet nor the strong transitions due to \FeII\  2382
and 2600 \AA, that would have fallen in the high SNR VIS region.
On the other hand, also our data show a tentative \CIV\ doublet at $z_{\rm abs}\simeq5.79$. Furthermore, we detected several strong low ionization lines at this redshift with a velocity profile fitted by three main components:  \SiII\ 1260, \OI\ 1302, \CII\ 1334, \FeII\  2344, 2382. 
Another metal system showing the same ionic transitions has been detected at  $z_{\rm abs}\simeq5.876$. 
A further paper will be devoted to the detailed analysis of these metal absorption systems (D'Odorico et al. in preparation). 

\subsection{CFHQS J1509-1749}

This object was discovered relatively recently (Willott et al. 2007) and
it is the least studied of our small sample. Our spectrum is, to our knowledge, 
the first to be reported in the literature at intermediate resolution. 
The SNR in the VIS region (redward of the
\Lya\ emission) varies between 40 and 15. In the NIR region, in
particular in the \CIV\ forest, SNR~$\sim 10-20$.

In the discovery paper, Willott and collaborators reported the detection
of two \MgII\ doublets at $z_{\rm abs} =3.266$ and 3.392 (the latter with
associated \MgI\ 2852 and \FeII\ 2344, 2382, 2586,
2600 \AA\ transitions). 
We confirm the presence of these two systems and add more transitions
to them (in particular \ZnII\ at $z_{\rm abs}=3.392$). Furthermore, we identify a
new \MgII\ doublet at $z_{\rm abs}=3.128$ with associated \FeII\ lines. 
In the VIS portion of the spectrum, we identify 5 \CIV\ doublets at
$z_{\rm abs}=4.611$, 4.642, 4.666, 4.792 and 4.816, while no \CIV\ is
detected at $z>5$ in the NIR spectrum.

\section{The \CIV\ mass density at $z\simeq4.5-5$}

The \CIV\ mass density, $\Omega_{\rm CIV}$, is a measure of the amount of metals present in the IGM at a given redshift (see e.g. D'Odorico et al. 2010 for the details of the computation).

\begin{table}
\begin{center}
\caption{$\Omega_{\rm CIV}$ for the systems in the total sample 
  selected in the reported column density intervals}
\label{omega}
\begin{tabular}{cc c c c c }
\hline  
$\log N($\CIV$)$ & $\langle z \rangle$ & $\Delta$ X & lines & $\Omega_{\rm CIV}$  & $\delta
\Omega$  \\ 
 & & & & ($\times 10^{-8}$) & ($\times 10^{-8}$)  \\
\hline
$13.0-15.0$ & $4.7$ &  10.08 & 32 & 2.6 &  0.6   \\
$13.8-15.0$ & $4.8$ & 10.08 & 10 & 1.7 & 0.6 \\
\hline
\end{tabular}
\end{center}
\end{table}

The value of $\Omega_{\rm CIV}$ in the redshift range $4.5 \lsim z \lsim 5.0$ was computed by Pettini et al. (2003) with the spectra of three high-$z$ quasars obtained with ESI at the Keck telescope. 
By adding the present sample to that of Pettini and collaborators, we increase by a factor $\sim3$ both the spanned redshift absorption path (from 3.5 to 10) and the number of detected lines\footnote{In the column density range $13 \le \log N_{\rm CIV} \le 15$} (from 11 o 32).
The computed $\Omega_{\rm CIV}$ for two ranges of \CIV\ column densities are reported in Table~\ref{omega}. They are in very good agreement with the original result by Pettini et al. (2003) corrected to the $\Lambda$CDM concordance cosmology and with the re-computation by Ryan-Weber et al. (2009)  to the column density range $13.8 \le \log N_{\rm CIV} \le 15$. 

This result put on a firmer ground the observed constant behaviour of $\Omega_{\rm CIV}$ in the redshift range $2.5 < z < 5$. More spectra are needed to compute a reliable value of $\Omega_{\rm CIV}$ beyond $z\sim5$: this is one of the final goals of our GTO programme. 
 
%
%
%
%

\acknowledgements
We are grateful to the team of people that projected, built and put
into operation such a wonderful instrument. In particular, our special
thank goes to Roberto Pallavicini, the Italian P.I. of X-shooter, who
did not have the time to see it working and producing scientific
results.        


\end{document}